\documentclass[english,aps,prbr,twocolumn,showpacs,superscriptaddress]
{revtex4-1}
\usepackage[T1]{fontenc}
\usepackage[latin9]{inputenc}
\usepackage{amsmath}
\usepackage{amssymb}
\usepackage{graphicx}
\usepackage{bm}
\usepackage{babel}
\usepackage{color}
\usepackage{hyperref}

\newcommand{\vk}{\mathbf{k}}
\newcommand{\vv}{\mathbf{v}}
\newcommand{\vq}{\mathbf{q}}
\newcommand{\vQ}{\mathbf{Q}}
\newcommand{\vl}{\mathbf{\Lambda}}
\newcommand{\eps}{\varepsilon}

\makeatletter

 \@ifundefined{textcolor}{}
 {
   \definecolor{BLACK}{gray}{0}
   \definecolor{WHITE}{gray}{1}
   \definecolor{RED}{rgb}{1,0,0}
   \definecolor{GREEN}{rgb}{0,1,0}
   \definecolor{DARKGREEN}{rgb}{0,0.5,0}
   \definecolor{BLUE}{rgb}{0,0,1}
   \definecolor{CYAN}{cmyk}{1,0,0,0}
   \definecolor{MAGENTA}{cmyk}{0,1,0,0}
   \definecolor{YELLOW}{cmyk}{0,0,1,0}
 }
\makeatother

\begin{document}

\title{Resistive anisotropy due to spin-fluctuation scattering\\ in the nematic
phase of iron pnictides}

\author{Maxim Breitkreiz}
\email{maxim.breitkreiz@tu-dresden.de}
\affiliation{Institute of Theoretical Physics, Technische Universit\"at Dresden,
01062 Dresden, Germany}

\author{P. M. R. Brydon}
\affiliation{Condensed Matter Theory Center, Department of Physics,
University of Maryland, College Park, USA 20742}

\author{Carsten Timm}
\email{carsten.timm@tu-dresden.de}
\affiliation{Institute of Theoretical Physics, Technische Universit\"at Dresden,
01062 Dresden, Germany}

\date{August 29, 2014}

\begin{abstract}
The large in-plane anisotropy of the resistivity is a hallmark of the nematic
state of the iron pnictides. Solving the Boltzmann transport equation, we show
that the prominent doping dependence as well as the large values of the
anisotropy can be well explained by momentum-dependent spin-fluctuation
scattering without assuming anisotropic impurity states. Due to the
forward-scattering corrections, the hot spots contribute to the resistive
anisotropy even in the case of strong spin fluctuations, which makes large
values of the anisotropy possible. The ellipticity of the electron pockets plays
an important role in explaining the dominance of positive values of the
anisotropy, i.e., larger resistivity in the direction with weaker spin
fluctuations, throughout the doping range.
\end{abstract}


\maketitle

\textit{Introduction.}
Currently, one of the most intensively discussed topics in the field of
high-$T_c$ superconductivity is the origin of the nematic phase
of the iron pnictides \cite{Fernandes2014, Davis2014}. The nematic phase
transition occurs at temperatures $T_s$ above or coinciding with the magnetic
ordering temperature $T_N$, at which a stripe antiferromagnetic
state with
ordering vector $\vQ_X=(\pi,0)$ (defining the
$x$-direction in this work) is established. The nematic phase found for
$T_N<T<T_s$ is characterized by a broken rotational symmetry between the
$x$ and $y$ directions in the absence of magnetic order.
Although one of its most obvious manifestations is the
orthorhombic distortion of the lattice, it is generally considered that
the nematic state arises from electronic correlations \cite{Chu2010}.
However, the precise mechanism is still under debate
\cite{Lv2011,Fernandes2012,Onari2012,Stanev2013,Liang2013}.

Another key experimental signature of the nematic phase
is the pronounced difference between the resistivities along
the $x$ and $y$ directions, $\Delta\rho\equiv(\rho_y-\rho_x)/\rho_x$
\cite{Chu2010, Ishida2013, Kuo2014, Blomberg2013, 
Ying2011}. Understanding the origin of the resistive anisotropy
should offer crucial insights into the origin of the nematicity. Two
scenarios are debated: (i) the 
scattering off anisotropic impurity states \cite{Allan2013, Ishida2013,
Inoue2012, Gastiasoro2013, Gastiasoro2014} and (ii) the
scattering off fluctuating
collective excitations with spectrum reflecting the underlying
nematicity \cite{Fernandes2011, Blomberg2013}.

The existing description of the resistive anisotropy due to spin
fluctuations \cite{Fernandes2011}, i.e., within scenario (ii), is restricted to
the limit of weak spin-fluctuation scattering compared to isotropic impurity
scattering, although the former is likely stronger than the latter, except 
at very low temperatures when the spin fluctuations are frozen 
out~\cite{Fang2009,Kasahara2010,Fanfarillo2012,Breitkreiz2013,Breitkreiz2014}.
Naturally, this limit is only compatible with small values
of $\Delta\rho$, since the dominant impurity part leads to isotropic
resistivity. Though in disagreement
with the huge positive anisotropy up to $\Delta\rho\approx0.5$ observed in
experiments on electron-doped samples \cite{Chu2010, Ishida2013}, the theory correctly
predicts negative $\Delta\rho$ for hole-doped samples \cite{Blomberg2013}.

Within scenario (i), the much larger $\Delta\rho$ in electron-doped
Ba(Fe$_{1-x}$Co$_x)_2$As$_2$ \cite{Chu2010} compared to
hole-doped Ba$_{1-x}$K$_x$Fe$_2$As$_2$ \cite{Blomberg2013,
Ying2011} is explained as a consequence of the stronger scattering off
Co-dopands placed \emph{within} the iron plane \cite{Ishida2013,
Allan2013,Gastiasoro2014}. The observed anisotropic impurity states
are all elongated in the $x$-direction, hence giving a larger
scattering cross-section in the
$y$-direction~\cite{Allan2013}. The negative $\Delta\rho$ measured for
hole-doped samples then arises due to
details of the band structure~\cite{Gastiasoro2014}.
The dependence of $\Delta\rho$ on the degree of disorder is
controversial: some experiments show, in agreement with scenario (i), a
reduction of $\Delta\rho$ upon sample annealing, which is supposed to lower the
degree of disorder \cite{Ishida2013}, while others report a much
weaker disorder dependence \cite{Kuo2014}.

In this work, we consider scenario (ii) with spin-fluc\-tu\-a\-tion scattering
of arbitrary strength. 
For spin-fluctuation and isotropic impurity scattering 
of comparable 
strength, we reproduce both
the small negative $\Delta\rho$ for hole-doped samples and the large positive
$\Delta\rho$ in electron-doped samples. We also show that the
reduction of $\Delta\rho$ in electron-doped samples upon 
annealing is consistent with the spin-fluctuation scenario. 
In a nutshell, our results follow from the role of the
spin-fluctuation scattering strength in controlling the size of the
Fermi-surface areas that contribute to the resistive anisotropy.

\begin{figure}[t]
\includegraphics[width=\columnwidth]{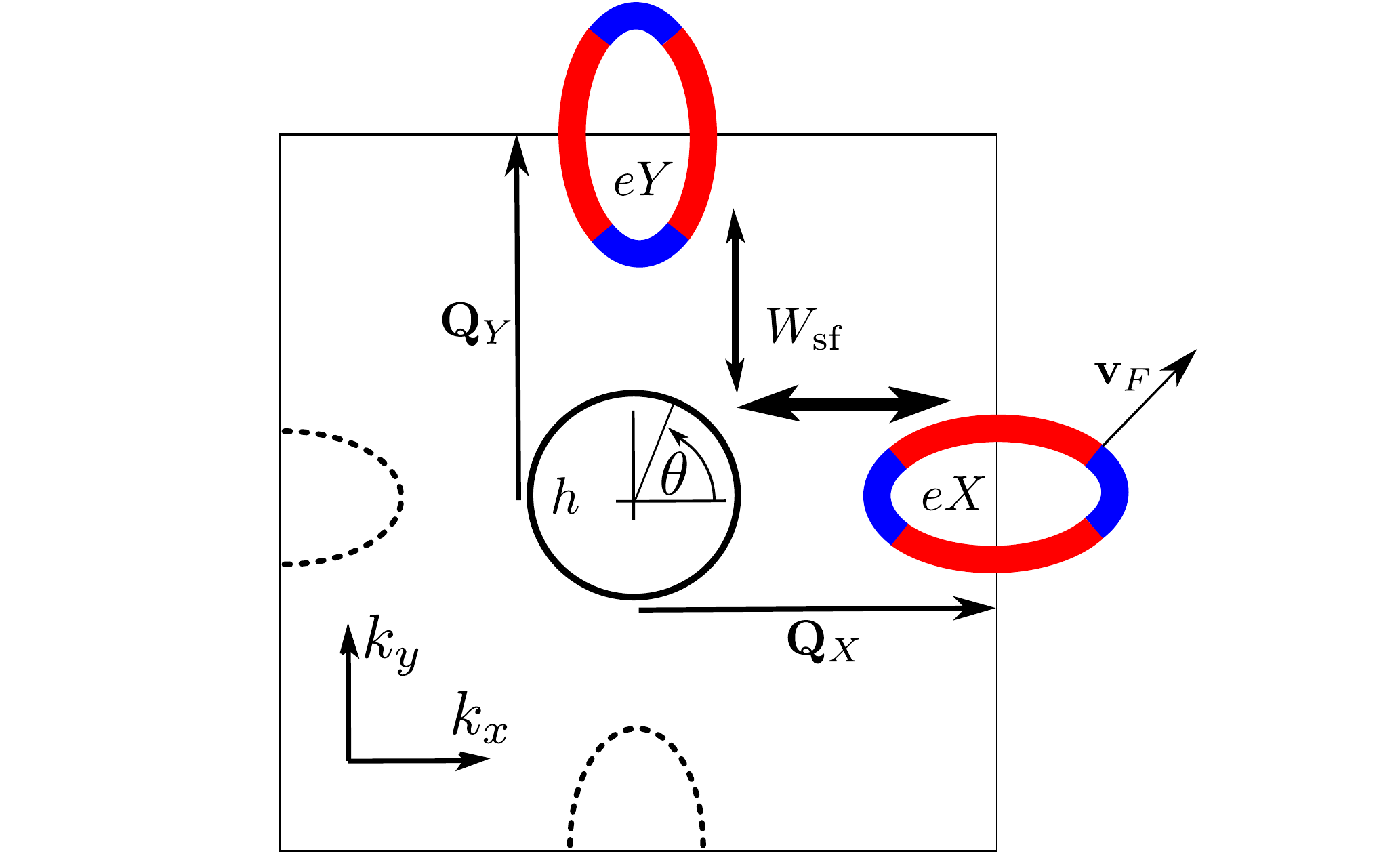}
\caption{(Color online) Hole (\textit{h}) and electron (\textit{eX} and
\textit{eY}) Fermi pockets of the two-band model.
In the nematic phase, scattering between \textit{h} and \textit{eX}
is stronger than between \textit{h} and \textit{eY}, as indicated by the
arrows marked $W_\mathrm{sf}$, giving rise to the resistive
anisotropy. As discussed in the main text, the electron pockets can be divided
into regions that contribute positively (red) or negatively (blue) to the
anisotropy, depending on the direction of the Fermi velocity.
States on each Fermi surface are parametrized by the angle $\theta$
to the $x$-axis with respect to the center of the pocket. }
\label{fig:figure1}
\end{figure}

\textit{Model and Method.}
We describe the band structure by an effective two-dimensional
model \cite{Fernandes2011,Fernandes2012,Brydon2011,Breitkreiz2014,Blomberg2013}
with a nearly circular hole Fermi
pocket at the center of the Brillouin zone and two elliptical electron
pockets \textit{eX} and \textit{eY} displaced by $\vQ_{X}=(\pi,0)$ and
$\vQ_{Y}=(0,\pi)$, respectively, where length is measured in units of
the iron-iron separation. 
We use the same dispersions as in Ref.\
\cite{Brydon2011} and fix the ellipticity of the electron pockets by
choosing $\xi_e=2$. The Fermi pockets are sketched in Fig.\
\ref{fig:figure1}. The sizes of the pockets depend on the doping level,
which is controlled by the electron filling $n$
\cite{Breitkreiz2014}.
The validity of the minimal model for the case of 122 pnictides 
has been discussed in the supplementary information for Ref.\ \cite{Blomberg2013}. 

To focus on the impact of the spin-fluctuation
scattering, in the following we neglect the distortion
of the Fermi pockets due to the splitting of the iron $d_{yz}$ and
$d_{xz}$ orbital levels \cite{Yi2011, Nakayama2014}. In the
Supplemental Material \cite{supp} we show that this
splitting gives rise
to an additional resistive anisotropy. By itself, this shows poor
agreement with experiment, however, and the effect of nematicity
in the spin-fluctuation scattering is the dominant mechanism
over a large parameter range.

We assume transport to be dominated by scattering off
spin fluctuations and isotropic impurities. The spin-fluctuation
scattering amplitude is determined by the imaginary part of the spin
susceptibility. We use a
phenomenological model for the susceptibility in the nematic phase
that has been employed for calculations in the impurity-dominated regime
\cite{Fernandes2011, Diallo2010, Inosov2010}. Following Ref.\
\cite{Breitkreiz2014}, we introduce a total elastic scattering rate
between states $|s,\theta\rangle$ on the Fermi pockets, parametrized by 
the pocket index $s$
and the angle $\theta$ (cf. Fig.\ \ref{fig:figure1}),
\begin{eqnarray}
\lefteqn{ W_{s\theta}^{s'\theta'} \equiv (1-\delta_{bb'})W_{\text{sf}}\, \alpha
} \nonumber \\
&& {}\times \int d\eps'\, \eps'\,
  \frac{\coth\frac{\eps'}{2k_BT}-\tanh\frac{\eps'}{2k_BT}}
  {\eps'^{2}+\omega_{\vq}^{2}}
  + W_{\text{imp}},
\label{eq:02}
\end{eqnarray}
where $\omega_{\vq} = \Gamma\,\big(\xi^{-2}\mp\phi+
q_x^2(1\pm\eta)+q_y^2(1\mp\eta)\big)$ with
$\vq=\vk(s,\theta,\eps_F)-\vk(s',\theta',\eps')$, where the wave vectors are
measured from the center of the corresponding Fermi pocket. Further, $b$
($b'$) is the band giving rise to the Fermi pocket $s$ ($s'$), $\phi$ is the
nematic order parameter, $\xi$ is the correlation length in the isotropic
phase, $\Gamma$ is the Landau damping parameter, and $\eta$ is the
in-plane anisotropy of the correlation length. The upper (lower)
sign corresponds to the scattering between the hole pocket and the electron
pocket \textit{eX} (\textit{eY}). $W_{\text{sf}}$ and $W_{\text{imp}}$ represent
the overall strengths of the scattering off spin fluctuations and impurities,
respectively, and the numerical factor $\alpha=10$
ensures that at the highest considered temperature (see below)
$W_{\text{sf}}/W_{\text{imp}}$
is of the same order as the inverse ratio of average lifetimes due to
scattering off spin fluctuations and impurities only,
$W_{\text{sf}}/W_{\text{imp}}\sim\tau_{\text{imp}}/\tau_{\text{sf}}$.

The susceptibility entering Eq.\ (\ref{eq:02})
is peaked at the nesting vectors $\vQ_X$ and $\vQ_Y$ for all dopings,
consistent with the observed stability of commensurate antiferromagnetic order
against doping. The
resulting scattering rate is therefore larger for
scattering wave vectors close to $\vQ_X$ or $\vQ_Y$. The strongest scattering is
found at the ``hot spots,'' i.e., the points on the Fermi pockets
connected by the nesting vectors. The position of the hot spots depends on
the doping level \cite{Brydon2011,Breitkreiz2014}.
In the nematic phase a finite
order parameter $\phi>0$ breaks the $C_4$ symmetry. This enhances the
peak at $\vQ_X$ in the susceptibility, leading to stronger scattering
between the hole pocket
\textit{h} and the electron pocket \textit{eX} than between the pockets
\textit{h} and \textit{eY}, as indicated in Fig.\ \ref{fig:figure1}.

\begin{figure*}[t]
\includegraphics[width=\textwidth]{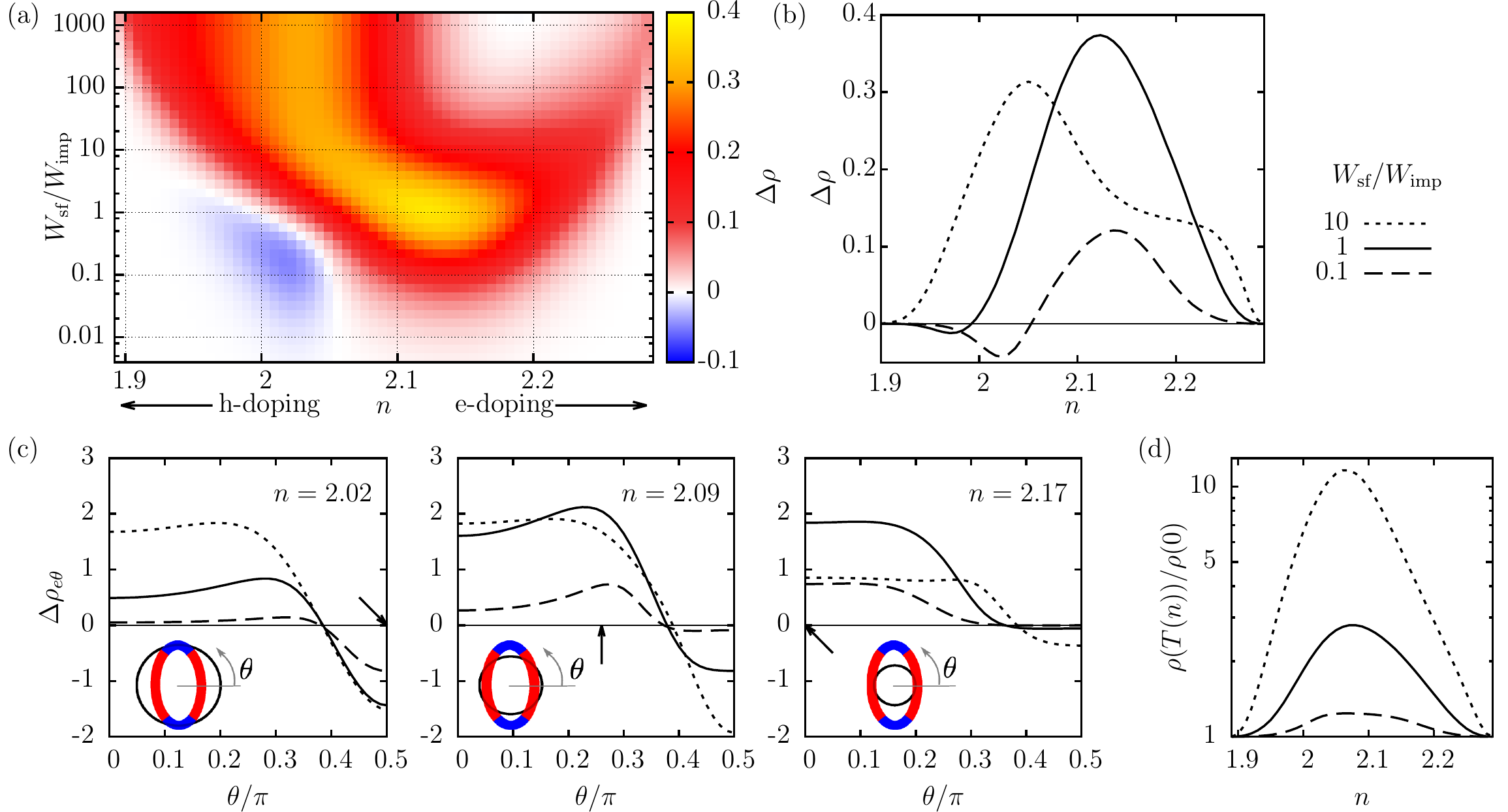}
\caption{(Color online) (a) Resistive anisotropy as a function of doping
(parametrized by $n$) and the
relative strengths of spin-fluctuation and impurity
scattering. (b) Resistive anisotropy as a function of doping for
$W_{\text{sf}}/W_{\text{imp}}=0.1$, $1$, and $10$. (c) Angle-resolved
contributions of the electron pockets to the resistive anisotropy as defined in
Eq.\ (\ref{eq:rce}). While for $W_{\text{sf}}/W_{\text{imp}}=0.1$ only regions
close to the hot spots (indicated by arrows) contribute, for increasing
$W_{\text{sf}}/W_{\text{imp}}$ the contributing regions grow.
(d) Ratio of averaged resistivities at the temperatures $T(n)$ considered in
(a)--(c) and at $T=0\,\mathrm{K}$.
We choose the parameters $\eta=0.5$,
$\Gamma=350$ meV, $\xi^{-2}=0.027$, and $\phi=0.017$.}
\label{fig:figure2}
\end{figure*}

We focus on the dependence of the resistive anisotropy
on doping (electron filling $n$)
and on the relative strengths of spin-fluctuation and
impurity scattering (controlled by $W_{\text{sf}}/W_{\text{imp}}$).
The explicit temperature $T$ in Eq.\ (\ref{eq:02}) controls the
energy available for spin excitations and thus additionally affects the
strength of spin-fluctuation scattering. In the relevant limit $k_BT\ll
\omega_\mathbf{q}$, this leads to the familiar $T^2$ dependence.
Since the nematic phase appears in a narrow temperature interval above
the N\'{e}el temperature $T_N(n)$, we choose the temperature 
$T(n) =T_N(n)  = T_0\,(1-[(n-2.09)/0.2]^2)$ with $T_0= \max[T_N(n)]
= 137\,\mathrm{K}$. This mimics the situation in 122 pnictides, where
the magnetic order is suppressed upon doping the parent compound, here taken to
correspond to $n=2.09$ \cite{Brydon2011}. Our results are qualitatively
insensitive to the specific form of $T(n)$.
Since the temperature tracks $T_N(n)$, it is reasonable to keep the
parameters $\xi$, $\phi$ , and $\Gamma$ fixed;
we have checked that the qualitative behavior does not depend on their
precise values.

We employ the non-equilibrium
Green-function formalism \cite{Rammer1986} in the Boltzmann
approximation, where
the linear-response distribution function at the Fermi
energy is determined by the vector
mean free paths $\vl_{s\theta}$ \cite{Mahan,SonTay} of the states
$|s,\theta\rangle$. 
The vector mean free path obeys the kinetic
equation \cite{Breitkreiz2014}
\begin{equation}
 \vl_{s\theta} = \tau_{s\theta}\, \vv_{s\theta}
  + \tau_{s\theta}\sum_{s'}\int \frac{d\theta'}{2\pi}\,
  N_{s'\theta'}\,W_{s\theta}^{s'\theta'}\,\vl_{s'\theta'},
  \label{eq:be}
\end{equation}
where $\mathbf{v}_{s\theta}
\equiv\hbar^{-1}\,\nabla_{\vk}\varepsilon_{b\vk}|_{s,\theta}$ is the velocity,
$N_{s\theta}=|d\vk_{s\theta}/d\theta|/\pi \hbar |\vv_{s\theta}|$ is the density of states, and
\begin{equation}
\tau_{s\theta}=\bigg(\frac{1}{2\pi}\sum_{s'}\int
  d\theta'\, N_{s'\theta'}\, W_{s\theta}^{s'\theta'}\bigg)^{\!-1},
\label{eq:lt}
\end{equation}
is the lifetime of the state $|s,\theta\rangle$. The first term on the
right-hand side of Eq.\ (\ref{eq:be}) represents the relaxation-time
approximation, while the second incorporates the forward-scattering
corrections.

The resistivity $\rho_i$ in the direction $i=x,y$ is determined by the vector
mean free path,
\begin{equation}
 \rho_i = \Big(e^2 \sum_{s} \int\frac{d\theta}{2\pi}\,
  N_{s\theta}\, v_{s\theta}^{i}\Lambda_{s\theta}^{i}\Big)^{\!-1}
  \equiv \Big(\sum_{s} \int\frac{d\theta}{2\pi}\,
    \sigma^{i}_{s\theta}\Big)^{\!-1},
  \label{eq:ra}
\end{equation}
where $\sigma^{i}_{s\theta}$ is the contribution of the state
$|s,\theta\rangle$ to the total conductivity
$\sigma^i=\sum_s\int\frac{d\theta}{2\pi}\,\sigma^i_{s\theta}$.
It is useful to resolve the resistive anisotropy in terms of
band and angular contributions,
\begin{equation}
\Delta\rho=\int\frac{d\theta}{2\pi}\,\big(\Delta\rho_{h\theta}+\Delta\rho_{
e\theta}\big),
\end{equation}
where the contributions from hole and electron pockets read, respectively,
\begin{eqnarray}
\Delta\rho_{h\theta} &\equiv&
  \frac{1}{2\sigma^y}\,
  \big(\sigma^x_{h,\theta}-\sigma^y_{h,\theta}
  +\sigma^x_{h,\theta+\pi/2}-\sigma^y_{h,\theta+\pi/2}\big),   \label{eq:rch} \\
\Delta\rho_{e\theta} &\equiv&
  \frac{1}{\sigma^y}\, \big(\sigma^x_{eY,\theta}-\sigma^y_{eY,\theta}
  +\sigma^x_{eX,\theta+\pi/2}-\sigma^y_{eX,\theta+\pi/2}\big).\quad
  \label{eq:rce}
\end{eqnarray}
In Eq.\ (\ref{eq:rch}), we consider the contributions from the
hole-pocket states $|h,\theta\rangle$ and $|h,\theta+\pi/2\rangle$
together, since only the joint contribution vanishes in the normal,
$C_4$-sym\-me\-tric phase. For the same reason, the states $|eY,\theta\rangle$
and $|eX,\theta+\pi/2\rangle$ are considered together in Eq.\ (\ref{eq:rce}).
According to the definition of $\Delta\rho_{e\theta}$, the contributions from
states close to the minor axis of the elliptical electron pockets are found at
$\theta\approx0$, while the contributions from states close to the major
axis are found at $\theta\approx\pi/2$.

\textit{Results.}
Figure \ref{fig:figure2} summarizes the results for the resistive anisotropy
obtained by solving Eq.\ (\ref{eq:be}) numerically \cite{Breitkreiz2014}.
In Fig. \ref{fig:figure2}(a) the resistive anisotropy
is plotted as a function of doping and the ratio
$W_{\text{sf}}/W_{\text{imp}}$, while in Fig.\ \ref{fig:figure2}(b) the doping
dependence is illustrated for three characteristic values of
$W_{\text{sf}}/W_{\text{imp}}$.
The contributions $\Delta\rho_{e\theta}$
from the electron pockets are found to dominate the
anisotropy, for which reason only these
contributions are shown in Fig.\ \ref{fig:figure2}(c). As
evident from Fig.\ \ref{fig:figure2}(c) and
illustrated in Fig.\ \ref{fig:figure1}, the electron pockets can be divided
into positively and negatively contributing parts, with the crossover located
roughly where the Fermi velocity points in the diagonal direction; the parts
close to the minor axis of the electron pockets contribute with positive sign,
while the parts close to the major axis contribute with negative sign.
This is because the conductivity of the electron pocket \textit{eY} is
larger than that of \textit{eX} due to the stronger scattering for the
latter.

\begin{figure}[t]
\includegraphics[width=\columnwidth]{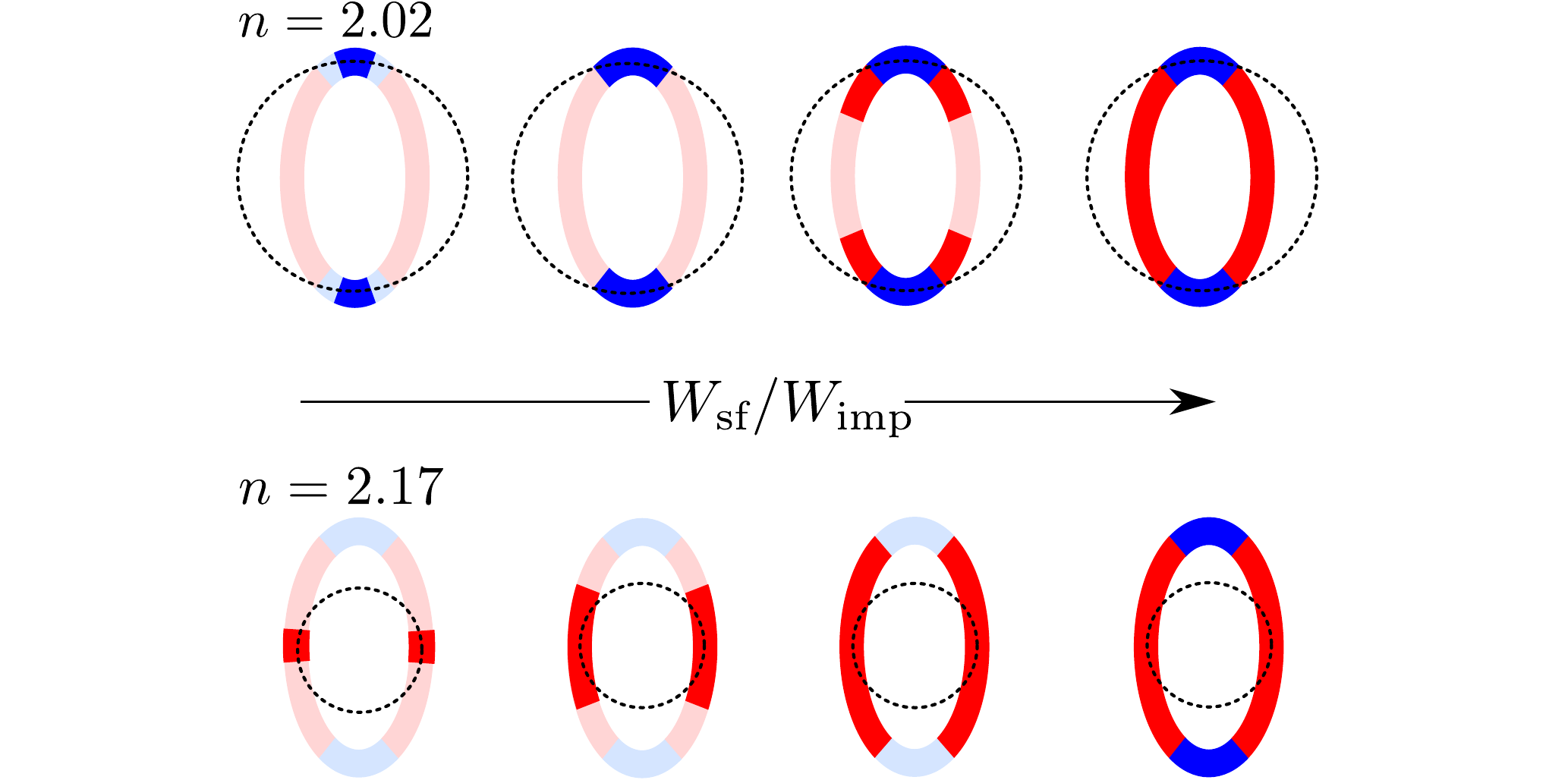}
\caption{(Color online) Increasing strength of spin-fluctuation scattering
extends the contributing regions of the electron pockets. Two characteristic
filling levels are considered, $n\approx2.02$ and $n\approx2.17$, with hot
spots at the major and the minor axis of the electron pockets, respectively.}
\label{fig:figure3}
\end{figure}

The total resistive anisotropy in Figs.\ \ref{fig:figure2}(a) and (b)
shows a strong doping dependence, which changes
qualitatively with $W_{\text{sf}}/W_{\text{imp}}$.
The angle-resolved plots in Fig.\ \ref{fig:figure2}(c) show that for increasing
$W_{\text{sf}}/W_{\text{imp}}$  the contributing regions of the electron
pockets expand. This is schematically illustrated in Fig.\ \ref{fig:figure3}.
For small  $W_{\text{sf}}/W_{\text{imp}}$, the resistive
anisotropy is dominated by regions close to the hot spots, whereas the
``cold'' regions, where spin-fluctuation scattering is weaker,
give small contributions. Since the electron pockets have
negatively and positively
contributing parts, the position of the hot spots determines
the sign of the resistive anisotropy. The negative (positive) extremum is found
for the filling $n\approx2.02$ ($n\approx2.17$), for which the hot spots lie
on the major (minor) axis of the electron pockets. The difference
between the positive and negative extrema is due to different velocities and
densities of states at the major and minor axes.

In the impurity-dominated limit,
$W_{\text{sf}}/W_{\text{imp}} \ll 1$, the anisotropy is very small as
impurity scattering is isotropic. With increasing
$W_{\text{sf}}/W_{\text{imp}}$, the contributing regions of the
electron pockets expand and the extrema of $\Delta\rho$ grow,
until the active region starts to include parts contributing with the opposite
sign. Upon further expansion, the positive and negative contributions
begin to partially compensate each other. Since the negatively
contributing regions are smaller, the negative extremum of
$\Delta\rho$ is suppressed at a smaller ratio
$W_{\text{sf}}/W_{\text{imp}}$ than the positive extremum. At
$W_{\text{sf}}/W_{\text{imp}}\approx1$ this results in a strong doping
asymmetry with small negative values on the hole-doped 
side and large
positive values on the electron-doped side. 

We emphasize that the result that the hot spots contribute to $\Delta\rho$
even for dominant spin-fluctuation scattering, as sketched in Fig.\
\ref{fig:figure3}, is
not obvious. Since in this limit the scattering at the hot spots is
much stronger than in the cold regions, one would
naively expect the hot spots to be short circuited by the cold
regions~\cite{Hlubina1995}, i.e.,
to be irrelevant for the transport, in which case $\Delta\rho$ would be
significantly smaller \cite{Fernandes2011,Blomberg2013}. However, as
we have
shown for the $C_4$-symmetric state of the pnicitides
\cite{Breitkreiz2014}, the short-circuiting is compensated by enhanced
forward-scattering corrections.

To compare the results to measurements, we have to
identify the relevant range of $W_{\text{sf}}/W_{\text{imp}}$.
In Fig.\ \ref{fig:figure2}(d), we plot the calculated ratio of the
averaged resistivity $\rho(T) \equiv (\rho_x+\rho_y)/2$ at
$T=T(n)$ and at $T=0\,\mathrm{K}$, where the spin excitations are
frozen out and the resistivity is due to impurity scattering alone,
which we assume to be temperature independent.
Ignoring for the moment that the system is antiferromagnetic at
$T=0\,\mathrm{K}$, we observe
that for $W_{\text{sf}}/W_{\text{imp}}=1$ and $W_{\text{sf}}/W_{\text{imp}}=10$
the resistivity ratios are comparable to those measured for as-grown and
annealed samples, respectively \cite{Ishida2013}.
The reduction of the density of states in the
antiferromagnetic phase should increase the $T=0\,\mathrm{K}$ resistivity,
however, and so our argument likely underestimates $W_{\text{sf}}/W_{\text{imp}}$.

For $W_{\text{sf}}/W_{\text{imp}}=1$, Figs.\ \ref{fig:figure2}(a) and
(b) show a large positive peak
with $\Delta\rho\approx0.4$ in electron-doped samples and a small
negative peak with $\Delta\rho\approx-0.01$ in hole-doped samples.
This is in good agreement with experimental observations \cite{Blomberg2013,Ishida2013,Chu2010}.
The results also show that in
electron-doped samples an increase of $W_{\text{sf}}/W_{\text{imp}}$
beyond about $1$ leads to a reduction of the peak value of
$\Delta\rho$. A reduction
of $\Delta\rho$ upon annealing was indeed observed in electron-doped
Ba(Fe$_{1-x}$Co$_x)_2$As$_2$ \cite{Ishida2013}, where this effect has been
taken as strong evidence that the resistive anisotropy mainly
stems from scattering at anisotropic impurity states. Our results show,
however, that such a reduction is also consistent with anisotropic
spin-fluctuation
scattering. For the
hole-doped samples, we predict an \emph{increase} in $\Delta\rho$ with
annealing if
$W_{\text{sf}}/W_{\text{imp}}\gtrsim1$, see Figs.\ \ref{fig:figure2}(a) and
(b), which to our knowledge has not been measured so far.

In the Supplemental Material \cite{supp}, we show that anisotropy due
to orbital splitting adds nearly additively to $\Delta\rho$, indicating
the robustness of the results against band details. This is in line with
the fact that
the main features of $\Delta\rho$ are explained by a mechanism that does not
rely on the details of the model.

\textit{Summary.}
We have studied the resistive anisotropy in the nematic state
of iron pnictides. We have considered a two-band model and assumed
scattering to be dominated by spin fluctuations and isotropic
impurities. The inclusion of forward-scattering
corrections is crucial for the correct description \cite{Breitkreiz2014}.
The obtained resistive anisotropy
$\Delta\rho$ shows good agreement with experimental results
for annealed and as-grown samples.
In particular, we have
shown that the twin puzzles of the doping asymmetry of
$\Delta\rho$
and the reduction of $\Delta\rho$ upon annealing
can be explained within the spin-fluctuation scenario. The qualitative behavior
is governed by the contributing regions on the elliptical
electron pockets, in particular their growth with increasing
spin-fluctuation strength.
Importantly, the hot spots contribute to $\Delta\rho$ even for strong
spin-fluctuation scattering, contrary to what was thought previously.
Since spin fluctuations are particularly strong at the hot spots, this
naturally leads to large anisotropies.

\textit{Acknowledgments.}
Financial support by the Deutsche Forschungsgemeinschaft through Research
Training Group GRK 1621 is gratefully acknowledged. The authors thank
B. M. Andersen, E. Babaev, M. N. Gastiasoro, P. J. Hirschfeld, D. Inosov,
and J. Schmiedt for useful discussions.

\clearpage
\onecolumngrid

\setcounter{figure}{0}
\setcounter{equation}{0}

\makeatletter 
\renewcommand{\thefigure}{S\@arabic\c@figure} 
\makeatother

\section*{Supplemental Material}

In the main text we calculate the resistive anisotropy due to
scattering off nematic spin fluctuations for a $C_4$-symmetric band structure.
The degeneracy of the iron $d_{yz}$ and $d_{xz}$ orbitals is lifted
in the nematic phase \cite{Yi2011, Nakayama2014}, however, lowering the
symmetry of the band structure to $C_2$. In this supplemental material we
consider the effect of this orthorhombic distortion in the band structure
on the resistive anisotropy.

\begin{figure}
\includegraphics[width=\textwidth]{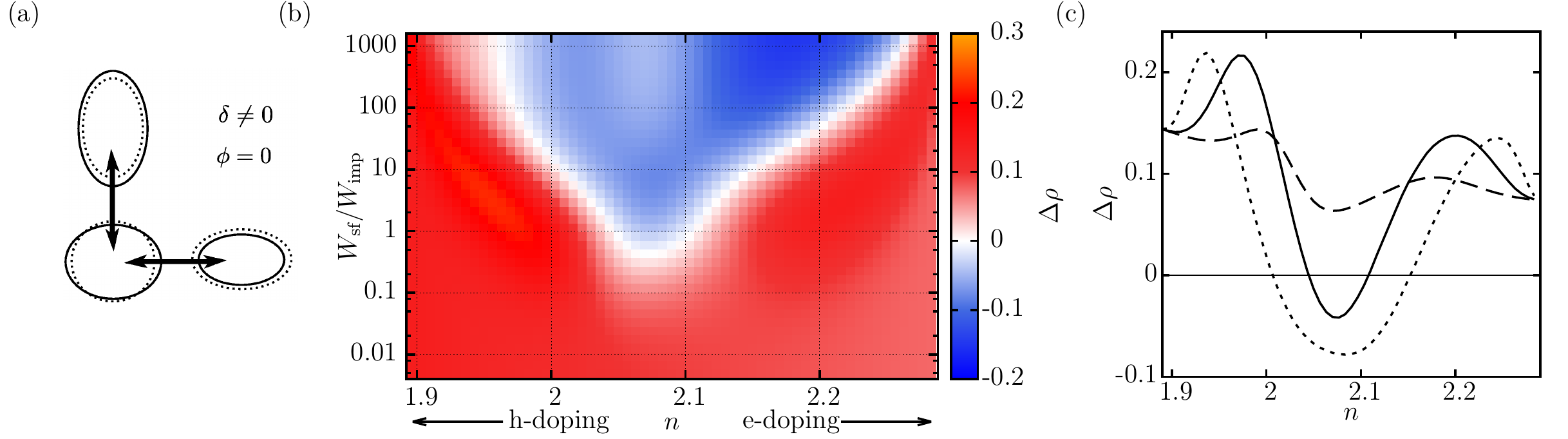}
\caption{(Color online) (a) Sketch of the Fermi pocket distortion and
the scattering strength between the hole and the electron
pockets. (b), (c) Resistive anisotropy in the presence of
orbital splitting ($\delta=0.03$) and a paramagnetic spin
susceptibility ($\phi=0$).\vspace{5ex}}
\label{fig:figureS1}
\end{figure}

The increased (decreased) iron-iron separation along the $x$ ($y$)
axis in the orthorhombic state decreases (increases) the onsite
energy of the iron $d_{xz}$ ($d_{yz}$) orbital. To model
the resulting changes in our band structure, we
follow Ref.\ \cite{Fernandes2014} and decrease the size of the
$eX$ pocket, increase the size of the $eY$ pocket, and elongate the
hole pocket along the $x$ direction, see Fig.~\ref{fig:figureS1}(a). This
distortion is motivated by the orbital
composition of the Fermi pockets~\cite{Graser2009}.
We implement the distortion by introducing a parameter
$\delta>0$ in the dispersion relations for the two bands $h$ and $e$:
\begin{eqnarray}
\eps_{h\vk}&=&\eps_{h}-\mu+2t_{h}\,\big[(1-\delta)\cos k_{x}+(1+\delta)\cos k_{y}\big],  \\
\eps_{e\vk}&=&\eps_{e}-\mu+t_{e,1}\cos k_{x}\,\cos k_{y}
{}- t_{e,2}\,\xi\,\big[(1+\delta)\cos k_{x}+(1-\delta)\cos k_{y}\big],
\label{eek}
\end{eqnarray}
where length is measured in units of the iron-iron separation. 
We choose a relatively large orthorhombic distortion of the band
structure with $\delta=0.03$, for which the relative difference of
the electron-pocket areas is about $21\%$. All other band parameters
are as in the main text.

For a nonzero orthorhombic distortion, the model displays a
resistive anisotropy $\Delta\rho$ even when the nematic
parameter in the susceptibility vanishes, $\phi=0$. 
We present results for this case in Fig.~\ref{fig:figureS1}. The
calculated $\Delta\rho$ is in rather poor agreement with experimental
findings: neither the minimum near optimal doping nor
the significant extent of negative values is observed. Note
that while the magnitude of $\Delta\rho$ scales with $\delta$, its
qualitative behavior does not change significantly.

\begin{figure}
\includegraphics[width=\textwidth]{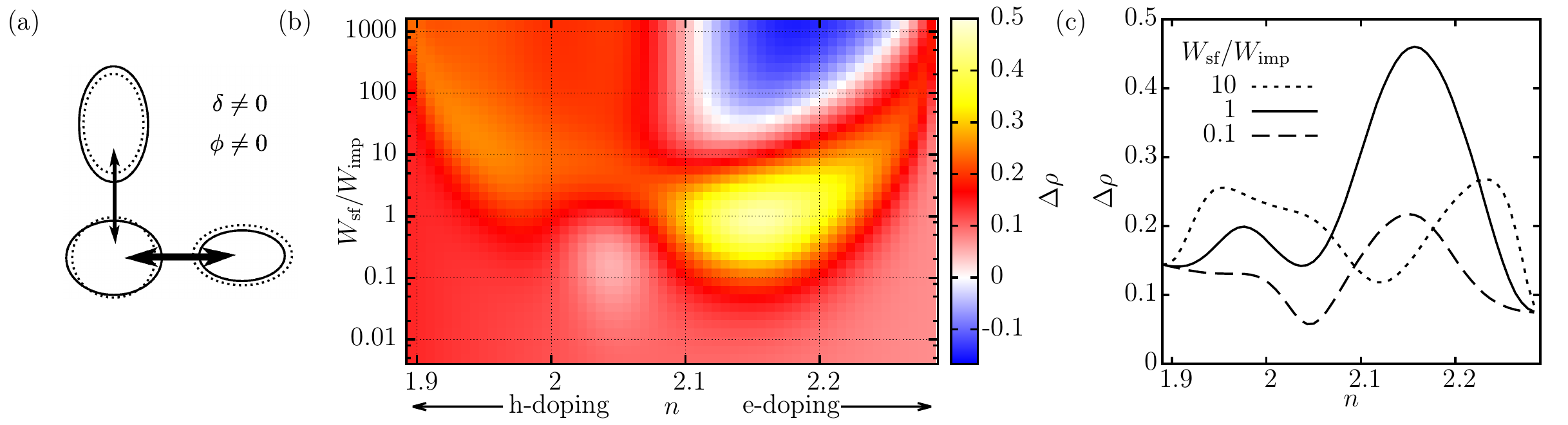}
\caption{(Color online) (a) Sketch of the Fermi pocket distortion and
the scattering strength between the hole and the electron
pockets. (b), (c) Resistive anisotropy in the presence of
orbital splitting ($\delta=0.03$) and nematic spin susceptibility
($\phi=0.017$).}
\label{fig:figureS2}
\end{figure}

Figure \ref{fig:figureS2} shows the result for the combined effect of
orbital splitting ($\delta=0.03$) and the nematicity in the spin
susceptibility ($\phi=0.017$). The effect of the two sources of
anisotropy appear to be additive and the characteristic signatures of the
nematic spin fluctuations are still conspicuous.
In particular, the large positive anisotropy in electron-doped samples and
the much smaller anisotropy in hole-doped samples for
$W_{\text{sf}}/W_{\text{imp}}\lesssim1$ is still present, as is the reduction
of the anisotropy in electron-doped samples for
$W_{\text{sf}}/W_{\text{imp}}\gtrsim1$. On the other hand, for
$W_{\text{sf}}/W_{\text{imp}}\gg1$, the weak contribution of the spin
fluctuations in the case of electron doping means that the resistive
anisotropy is controlled by the distortion of the band structure
and becomes negative, as in Fig.\ \ref{fig:figureS1}.

In summary, the effect of orbital splitting alone cannot account
for the observed resistive anisotropy. Better agreement might be
achieved for a more sophisticated model of the band structure,
although this would be at the expense of fine tuning. In contrast,
including the nematicity in the spin fluctuation spectrum gives much better
agreement with experimental results, is robust against the
distortion of the band structure, and dominates the contribution of
the distorted band structure to the resistive anisotropy over a
large parameter range.

\newpage

~

\end{document}